\documentclass[graybox]{svmult}

\usepackage{mathptmx}       
\usepackage{helvet}         
\usepackage{courier}        
\usepackage{type1cm}        
\usepackage{makeidx}         
\usepackage{graphicx}        
\usepackage{multicol}        
\usepackage[bottom]{footmisc}

\makeindex             

\begin{document}

\title*{The [$\alpha$/Fe] Ratios in Dwarf Galaxies: Evidence for a 
Non-universal Stellar Initial Mass Function? }
\titlerunning{[$\alpha$/Fe] Ratios in Dwarf Galaxies and the variable IMF}
\author{Simone Recchi, Francesco Calura and Pavel Kroupa}
\institute{Simone Recchi \at Institute of Astronomy, Vienna University, 
\email{simone.recchi@univie.ac.at}
\and Francesco Calura \at Jeremiah Horrocks Institute, 
University of Central Lancashire \email{fcalura@uclan.ac.uk}
\and Pavel Kroupa \at Argelander Institute for Astronomy, 
Bonn University \email{pavel@astro.uni-bonn.de}}
%
%
\maketitle

\abstract{It is well established that the [$\alpha$/Fe] ratios in 
elliptical galaxies increase with galaxy mass.  This relation holds
also for early-type dwarf galaxies, although it seems to steepen at
low masses.  The [$\alpha$/Fe] vs. mass relation can be explained
assuming that smaller galaxies form over longer timescales
(downsizing), allowing a larger amount of Fe (mostly produced by
long-living Type Ia Supernovae) to be released and incorporated into
newly forming stars.  Another way to obtain the same result is by
using a flatter initial mass function (IMF) in large galaxies,
increasing in this way the number of Type II Supernovae and therefore
the production rate of $\alpha$-elements.  The integrated galactic
initial mass function (IGIMF) theory predicts that the higher the star
formation rate, the flatter the IMF.  We have checked, by means of
semi-analytical calculations, that the IGIMF theory, combined with the
downsizing effect (i.e. the shorter duration of the star formation in
larger galaxies), well reproduces the observed [$\alpha$/Fe] vs. mass
relation.  In particular, we show a steepening of this relation in
dwarf galaxies, in accordance with the available observations. }

\section{Introduction}
\label{sec:1}
The integrated galactic initial mass function (IGIMF) theory
\cite{wk05} is based on the following 3 assumptions:
\begin{itemize}
\item most stars in galaxies form in star clusters (SCs). Within each 
SC, the IMF can be approximated by $\xi(m)\propto m^{-\alpha}$, with
$\alpha=1.3$ for $m<0.5$ M$_\odot$ and $\alpha=2.35$ for $0.5$
M$_\odot<m<m_{max}$. The value of $m_{max}$ depends on the mass of the 
SC (the larger the mass, the higher the probability of forming massive
stars).
\item Also the SCs are distributed according to a power law, 
$\xi_{ecl}\propto {M_{ecl}}^{-\beta}$, where $M_{ecl}$ is
the mass of the SC.  Observations suggest $\beta$ to be about 
2 \cite{ll03}.  
\item The maximum possible mass of a SC increases with the star 
formation rate (SFR) of the galaxy \cite{lr00}.  The IGIMF in galaxies
depends thus on the SFR. Galaxies with high SFRs contain larger
clusters and, consequently, a larger fraction of massive stars. The
IMF is therefore flatter than in galaxies with low SFRs. 
\end{itemize}

\section{Aims and assumptions}
\label{sec:2}
We study how the [$\alpha$/Fe] ratios in galaxies are affected by the
steepening of the IGIMF with decreasing SFR.  Our aim is to reproduce
the correlation between [$\alpha$/Fe] and mass (or velocity dispersion
$\sigma$) observed in early-type galaxies. We assume for simplicity
the SFR to be constant over a period of time $\Delta t$, which
increases with decreasing galaxy mass, in compliance with downsizing
(\cite{thom05}; see Fig. \ref{fig:2}, right panel, dashed line). The
Type Ia (main producers of Fe) and Type II (main producers of
$\alpha$-elements) Supernova rates can be thus calculated
analytically, within the IGIMF framework.

%
\begin{figure}[t]
\centering\includegraphics[width=7cm]{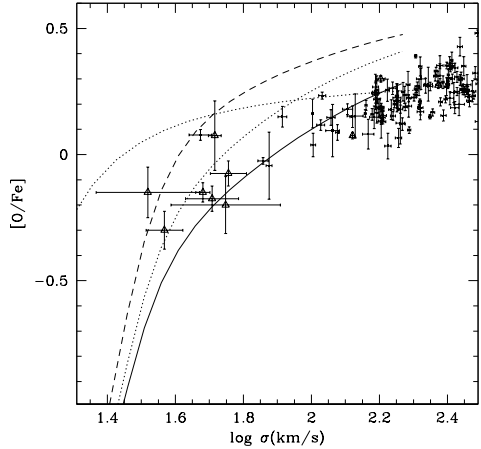}
%
\caption{[O/Fe] vs. $\sigma$ for models with a constant SFR over a 
period of time $\Delta t$ (see text).  Different values of $\beta$
(slope of the SC mass distribution function) are considered: 2.35
(heavy solid line), 2 (heavy dotted line) and 1 (heavy dashed line).
Also plotted is a model with $\beta=2$ and $\Delta t=1$ Gyr,
irrespective of the galaxy mass (light dotted line).  Observational
data are from \cite{thom05} (filled squares) and from \cite{sn08}
(open triangles).}
\label{fig:1}       
\end{figure}

\section{Results and conclusions}
\label{sec:3}
We calculate both mass-averaged and luminosity-averaged [$\alpha$/Fe]
ratios in early-type galaxies of different masses (which we convert to
$\sigma$ through the Faber-Jackson relation). In Fig. \ref{fig:1} we
compare the resulting [O/Fe] (without fine tuning of the parameters)
as a function of the slope of the SC mass distribution function
$\beta$ (heavy lines) with the available observations.  The evolution
of [Mg/Fe] is very similar to the [O/Fe] evolution and we do not
report it here.  We reproduce qualitatively the increase of
[$\alpha$/Fe] with $\sigma$. This result is partially due to the
downsizing (in low-mass galaxies the SFR lasts longer and the SNeIa
have more time to pollute the galaxy).  To demonstrate that, we plot a
model with $\beta=2$ and with constant $\Delta t=1$ Gyr, irrespective
of the galaxy mass (light dotted line).  This model turns out to be
too flat and does not reproduce well the data, therefore the downsizing 
is a necessary ingredient to reproduce the [$\alpha$/Fe] vs. $\sigma$ 
relation.

However, because of the steeper IMF in dwarf galaxies (resulting in a
lower SNII rate), our theoretical curves bend down at low
$\sigma$. This result is in agreement with the available observations.

\begin{figure}
\includegraphics[width=.5\textwidth]{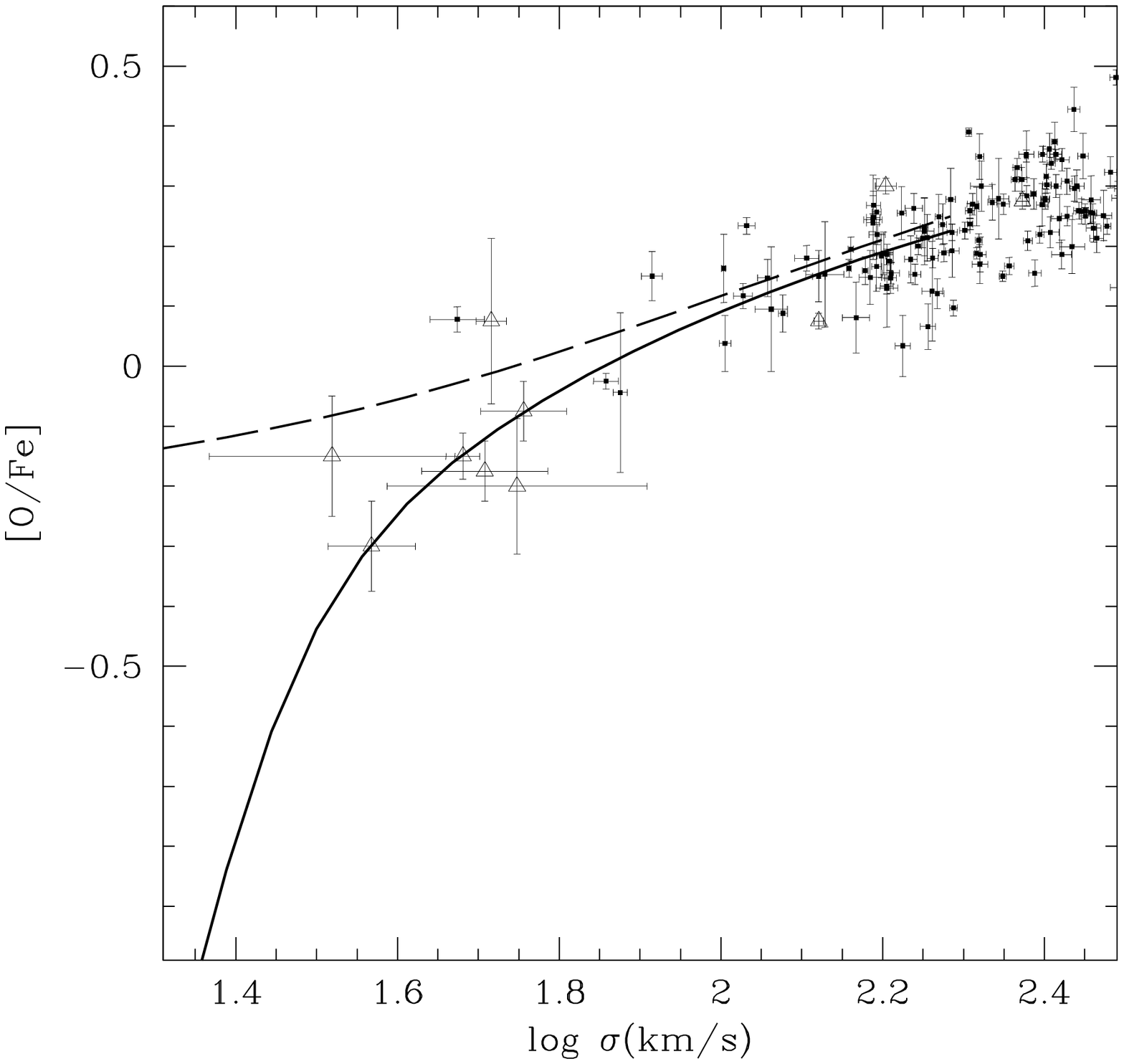}
\includegraphics[width=.5\textwidth]{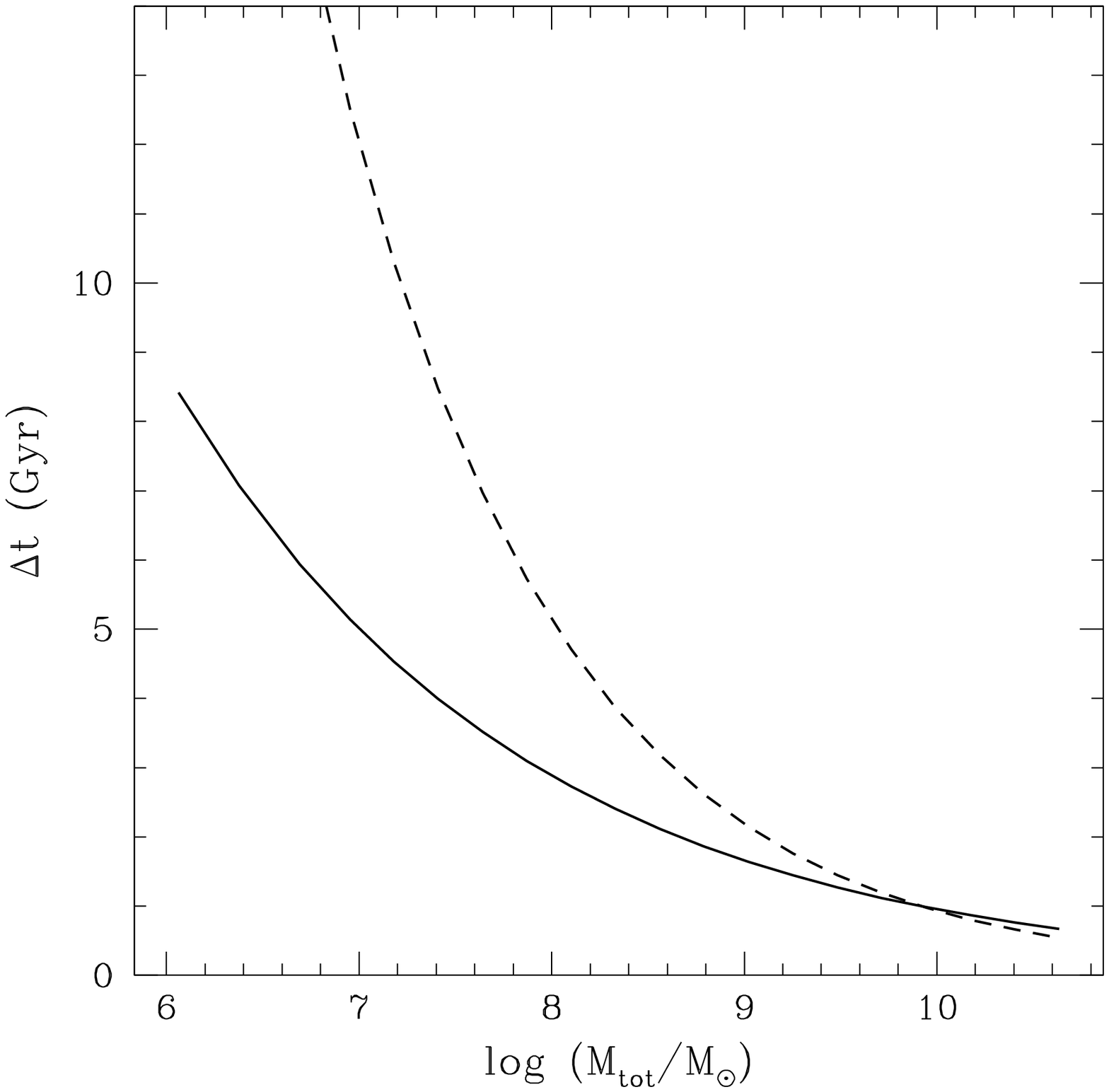}
\caption
{ (Left panel) [O/Fe] vs. $\sigma$ for our best fit IGIMF model (solid
line) and for our best fit model with constant IMF (long-dashed line).
(Right panel) $\Delta t$ vs. luminous mass relation that has been
employed in the best fit model in the left panel (solid line) and
$\Delta t$--mass relation obtained by \cite{thom05} (dashed line).  }
\label{fig:2}
\end{figure}

In order to best fit the available data, we allow a variation in the
SNIa parameters and, most importantly, in the $\Delta t$-luminous mass
relation. Our best fit is shown in Fig. \ref{fig:2}, left panel (solid
line) and is based on the relation shown in Fig. \ref{fig:2}, right
panel (solid line).  From this plot we can notice that the downsizing
effect (namely the shorter duration of the star formation in larger
galaxies) is milder than in \cite{thom05}, in the sense that the
$\Delta t$ for large galaxies is slightly larger than the timescale
calculated by \cite{thom05}.  Also shown in Fig. \ref{fig:2}, left
panel is the best fit [$\alpha$/Fe] vs. $\sigma$ obtained with a
constant (i.e. not SFR-dependent) IMF (long-dashed line), obtained
using the same modified $\Delta t$-luminous mass relation employed for
the best fit IGIMF model. This curve reproduces well the data points
at high $\sigma$ but is less accurate at low $\sigma$ because the
[$\alpha$/Fe] does not steepen, as the available observations seem to
indicate. It is also to note that the best fit has been obtained with
$\beta$=2.35. More details about these calculations in \cite{r09}.  It
is clear that more observations of the [$\alpha$/Fe] ratios in dwarf
galaxies are needed to test the validity of our results (see for
instance the contribution of A. Rys in this volume).

The main conclusions of this work can be summarized as follows:
\begin{itemize}
\item Models in which the IGIMF theory is implemented naturally 
reproduce an increasing trend of [$\alpha$/Fe] with $\sigma$, as 
observed in early-type galaxies.
\item These models show (at variance with constant IMF models) a 
steepening of the  [$\alpha$/Fe] vs. $\sigma$ relation for small 
galaxies, as the observations indicate.
\item In order to best fit the observations, the downsizing effect 
(namely the shorter duration of the star formation in larger galaxies) 
has to be milder than previously thought.
\item The best results are obtained for a star cluster mass 
function $\xi_{\rm ecl}\propto M_{\rm ecl}^{-2.35}$.
\end{itemize}

%

%
%
%

\end{document}